# Resistive Switching in Cr doped $SrTiO_3$: An X-ray Absorption Spectroscopy study

B. P. Andreasson[1*], M. Janousch[1], U. Staub[1], G. I. Meijer[2] and B. Delley[1]

[1]Swiss Light Source, Paul Scherrer Institut, 5232 Villigen PSI, Switzerland

[2]IBM Research, Zurich Research Laboratory, 8803 Rüschlikon, Switzerland

[*]*corresponding author, email: pererik.andreasson@psi.ch*

## *Abstract*

X-ray absorption spectroscopy was used to study the microscopic origin of conductance and resistive switching in chromium doped strontium titanate ($Cr:SrTiO_3$). Differences in the x-ray absorption near edge spectroscopy (XANES) at the Cr *K*-edge indicate that the valence of Cr changes from 3+ to 4+ underneath the anode of our sample device after the application of an electric field. Spatially resolved x-ray fluorescence microscopy (μ-XRF) maps show that the $Cr^{4+}$ region retracts from the anode-$Cr:SrTiO_3$ interface after a conducting state has been achieved. This interface region is studied with extended x-ray absorption fine structure (EXAFS) and the results are compared with structural parameters obtained from density functional theory (DFT) calculations. They confirm that oxygen vacancies which are localized at the octahedron with a Cr at its center are introduced at the interface. It is proposed that the switching state is not due to a valence change of chromium but caused by changes of oxygen vacancies at the interface.

## *Introduction*

The current research interest in new materials and processes for non-volatile memory applications includes magnetic random-access-memory[1], phase-change memory[2] and resistance-change memory based on transition-metal oxides[3]. It has been shown that a number of different transition-metal oxides exhibit resistive switching, e.g. chromium doped $SrTiO_3$[4], titanium dioxide[5], cuprates[6] and manganites[7]. In some cases the material is initially insulating and has to be electrically formed by an application of a strong electric field, to induce a conducting state at which resistive switching occurs. Several models have been proposed to explain the resistive switching in such transition-metal oxides: e.g. conduction based on crystalline defects due to the applied electric field[8], phase separation of non-percolative metallic domains[9] and a Mott metal-insulator transition at the interface [10]. However, the underlying microscopic mechanism of the resistive switching is still debated.

$SrTiO_3$ crystallizes in a simple cubic perovskite lattice. $SrTiO_3$ is a band insulator, for which a small amount of carrier doping, approximately $10^{18}$ $e$/cm$^3$, leads to a conducting state. This can be achieved by e.g. introducing oxygen vacancies or substituting Ti with Nb[11]. For chromium doped $SrTiO_3$, $Cr^{3+}$ replaces the $Ti^{4+}$ as the center ion of the oxygen octahedron. As $Ti^{4+}$ is in a $3d^0$ (s=0) state and $Cr^{3+}$ ($3d^3$) and $Cr^{4+}$ ($3d^4$) are both stable oxidation states, the role of the Cr valence in conductance and the switching behavior needs to be addressed.

Recently, Cr *K*-edge x-ray absorption near edge spectroscopy (XANES) on chromium doped $SrTiO_3$ found that the chromium valence is 4+ underneath the cathode, and 3+ elsewhere[12]. More recently, a XANES study with much improved lateral resolution concluded that oxygen vacancies are introduced at the octahedron surrounding the Cr in the conducting region between the two electrodes[13]. These vacancies lead to a distinct enhancement of the absorption in the Cr pre-edge region, largest at the electrode-crystal interfaces. It has been proposed that these defects are responsible for the conductance and the switching.

Here, we present extended x-ray absorption fine structure (EXAFS) data, which is used to probe the proposed structural change of the oxygen octahedron with Cr as center absorbing atom. In addition, to resolve the role played by the chromium valence we monitored the change of the chromium *K*-edge XANES during the electrical forming.

## *Experimental*

Single crystals of 0.2 mol% Cr-doped $SrTiO_3$ are grown in a $N_2/O_2$ atmosphere by floating-zone melting, as in reference[12]. Four pairs of platinum electrodes are patterned using standard photo lithography in planar geometry along the [010] axis on the [001] surface of the crystals. Their sizes are 400×50 $\mu m^2$, 100 nm thick and separated by a 50 μm gap. The electrodes are contacted by 100 μm thick indium wires, which are connected to an integrated circuit carrier.

In the initial state the crystal is insulating. Applying a strong electric field (80 kV/cm) over the electrodes for several minutes induces a metallic path. This process will be called electrical forming in the rest of the proceeding. In figure 1(a) the measured current as a function of time for the first part of the electrical forming is plotted. After forming to 1 μA, the applied electric field is lowered while the current limit is successively increased until a current of 10 mA is flowing at an applied voltage of approximately 20V. After the electrical forming, I-V scans between -20 V and 20 V are performed, see figure 1(b). The observed hysteresis curve confirms the existence of the two resistance states as shown by their different slopes at low voltage. Initially, the crystals are transparent and light yellow in color. After the electrical forming, a path is seen between the electrodes in the visible light micrograph, see figure 2(a).

The x-ray absorption measurements were performed at the LUCIA beamline[14] of the Swiss Light Source at the Paul Scherrer Institut. The beam size was approximately 3×3 $\mu m^2$, probing a few micrometer sample depth at the Cr *K*-edge (~6000 eV). The sample surface made an angle of approximately 45° with respect to the incoming beam. The absorption was measured by recording the fluorescence yield of the Cr $K_\alpha$ line with a silicon drift detector.

### *Results*

Figure 2(b) shows three typical Cr *K*-edge XANES spectra taken on the sample. The solid line is the reference taken prior to electrical forming (dot-dashed square, figure 2(a)), the circles show data taken in the cathode (or anode) interface after electrical forming (solid squares, figure 2(a)) and the squares show data taken beneath the anode where $Cr^{4+}$ is detected (dotted square, figure 2(a)).

To monitor the change in chromium valence during the electrical forming, we chose an x-ray energy were the contrast between the $Cr^{3+}$ and $Cr^{4+}$ shows the largest difference. At this energy (6007.3 eV) the sample is spatially scanned in the beam over a 100×150μm$^2$ large area centered at the gap between the electrodes. The results of these measurements are seen in figure 2(c) and (d): (c) is taken in the initial part with voltage applied but no current flowing yet and (d) is taken after the sample was formed to conduct 10 mA. The brighter area represents $Cr^{3+}$, the darker area corresponds to $Cr^{4+}$. It can be seen that initially the $Cr^{4+}$ region is restricted to the anode and retracts from the area where the electrical path connects to the electrode in the conducting state.

To test the hypothesis of oxygen vacancies located at the Cr atom in the electrode-interface region reported by previous XANES measurements[13], Cr *K*-edge (EXAFS) data were taken. The data were normalized and the background was removed by standard EXAFS procedures and transformed to momentum space. Figure 3 shows the $k^2$ weighted EXAFS function $\chi(k)$. The solid line shows data from the unformed reference (dot-dashed square figure 2(a)) and the circles connected with a dashed line show data taken at the cathode interface (solid square figure 2(a)) of a fully formed sample. The location on the cathode was chosen where the XANES data showed the highest deviation from the unformed reference in the pre-edge region[13].

The EXAFS analysis was performed using the graphical user interface ARTEMIS[15] which uses the computer code IFEFFIT[16]. Theoretical spectra were obtained by the FEFF6[17] code. Calculations were based on an undistorted crystal structure of $SrTiO_3$ with space group Pm-3m, lattice parameter 3.95 Å. Single and multiple scattering paths were generated up to a path length of 4 Å. Potentials were calculated with 87 atoms where the central titanium atom was replaced by a chromium atom. The theoretical spectra were fitted in Fourier space to the experimental ones up to 4 Å in a $k$-range from 4 $Å^{-1}$ to 9 $Å^{-1}$. The Sr and Ti shells as well as multiple scattering paths were included for reliability of the fits. For the EXAFS reference spectrum of an unformed crystal, a full oxygen octahedron with six nearest neighbors was used. The free parameters were the passive electron reduction factor $S_0^2$ for oxygen, change in average nearest neighbor distance $\Delta R$, Debye-Waller factor $\sigma_O^2$ for the oxygen and the shift in energy $\Delta E$. The resulting fit and data of unformed $SrTiO_3$ can be seen in figure 4(a) as solid line and squares respectively. The parameters obtained from the fit of the first shell were $S_0^2 \approx 0.92$, $\Delta R \approx -0.034$Å, $\Delta E \approx -4$eV and $\sigma_O^2 \approx 0{,}004$Å$^2$. To consistently fit the EXAFS data of the interface region, $\Delta E$ and $\sigma_O^2$ were kept fixed to the values from the fitting of the reference. Since in this case, $N$ and $S_0^2$ are strongly correlated, the number of nearest neighbors was set to $N$=5 (corresponding to one missing oxygen in the octahedron). The resulting fit and data of fully formed $SrTiO_3$ can be seen in figure 4(b) as solid line and circles respectively. The parameters obtained from this fit were $S_0^2 \approx 0.92$, which shows that the reduction of $N$ is meaningful, and $\Delta R \approx -0.016$Å.

## *Discussion*

In the initial state the formal valence of chromium is homogeneously 3+ throughout the sample. When applying a voltage, at first no current is flowing but the chromium valence changes to 4+ beneath the anode. When a small current is flowing, the valence changes back to $Cr^{3+}$ beneath the part of the anode that has contact with the conducting part of the crystal. Therefore, in the beginning the applied field removes electrons from the anode and may polarize it. A flowing current leads to a close to zero effective electrical potential which decreases the Cr valence back to its ground state as $Cr^{3+}$. As the path was found to be metallic[12], the $Cr^{4+}$ is therefore not likely playing a role in the switching process. The chromium valence change observed previously[12] is probably an initial phenomenon due to the strong electrical potential in the forming process.

The amplitude change seen in the $k^2$ weighted $\chi(k)$ can originate from fewer nearest neighbors which is corroborated with our fit with $N$=5 instead of $N$=6 as for the unformed reference. The oxygen defect in the octahedron with Cr at its center is consistent with our previous Cr *K*-edge XANES interpretation[13]. It is also consistent with publications from other groups suggesting that oxygen vacancies or defects are important for the switching process[18, 19].

From the EXAFS data we obtain a Cr-O bond length of 1.95 Å and 1.97 Å for the unformed and interface region, respectively. This result compares well to the one obtained from density functional theory (DFT) calculations, where the relaxed mean distances between the chromium and the oxygen ions are calculated to be 1.91 Å and 1.93 Å for the regular octahedron and with one oxygen missing, respectively. Experiment and theory give an increase in the Cr-O bond length when an oxygen atom is removed from the octahedron.

## *Conclusions*

The EXAFS measurements in the cathode interface show that at the Cr the number of nearest neighbors is reduced and the Cr-O distance slightly increased. These changes are exclusively affecting the Cr as no observable changes in the Ti *K*-edge XANES were found, and confirm the importance of oxygen defects for conductivity and switching. From the μ-XRF measurements it is concluded that the formal valence of the chromium does not play a crucial role in the switching process.

### *Acknowledgements*

This work was performed at the Swiss Light Source, Paul Scherrer Institut, Villigen, Switzerland. For outstanding technical support at the LUCIA beamline, R. Wetter is acknowledged.

*Figure 1*

(a) Time dependence of the current through the sample, at a constant applied voltage (400 V). (b) I-V characteristics of a fully formed sample which illustrates the two resistance states by the different slopes in the low voltage region.

*Figure 2*

(a) Visible light image of the device. The conducing path is seen as the darker region between the anode and the cathode. (b) Chromium *K*-edge XANES spectra taken on the unformed reference (solid line, dash-dotted square in (a)), beneath the anode (squares, dotted square in (a)) and in the electrode-$SrTiO_3$ interfaces (circles, solid squares in (a)). (c) The μ-XRF map taken at 6007.3 eV after voltage has been applied but without current flowing and (d) the μ-XRF map taken at 6007.3 eV on the fully formed sample. In these maps the darker regions represent $Cr^{4+}$ and the brighter regions correspond to $Cr^{3+}$.

*Figure 3*

The $k^2$ weighted EXAFS function $\chi(k)$ as extracted from the data. Solid line represents data taken at the unformed reference and circles connected with a dashed line are taken in the cathode-interface.

*Figure 4*

Fourier transform of the $k^2$ weighted EXAFS function: (a) the unformed reference position, showing data as squares and fit as solid line and (b) the electrode interface, showing data as circles and fit as solid line.

Figure 1

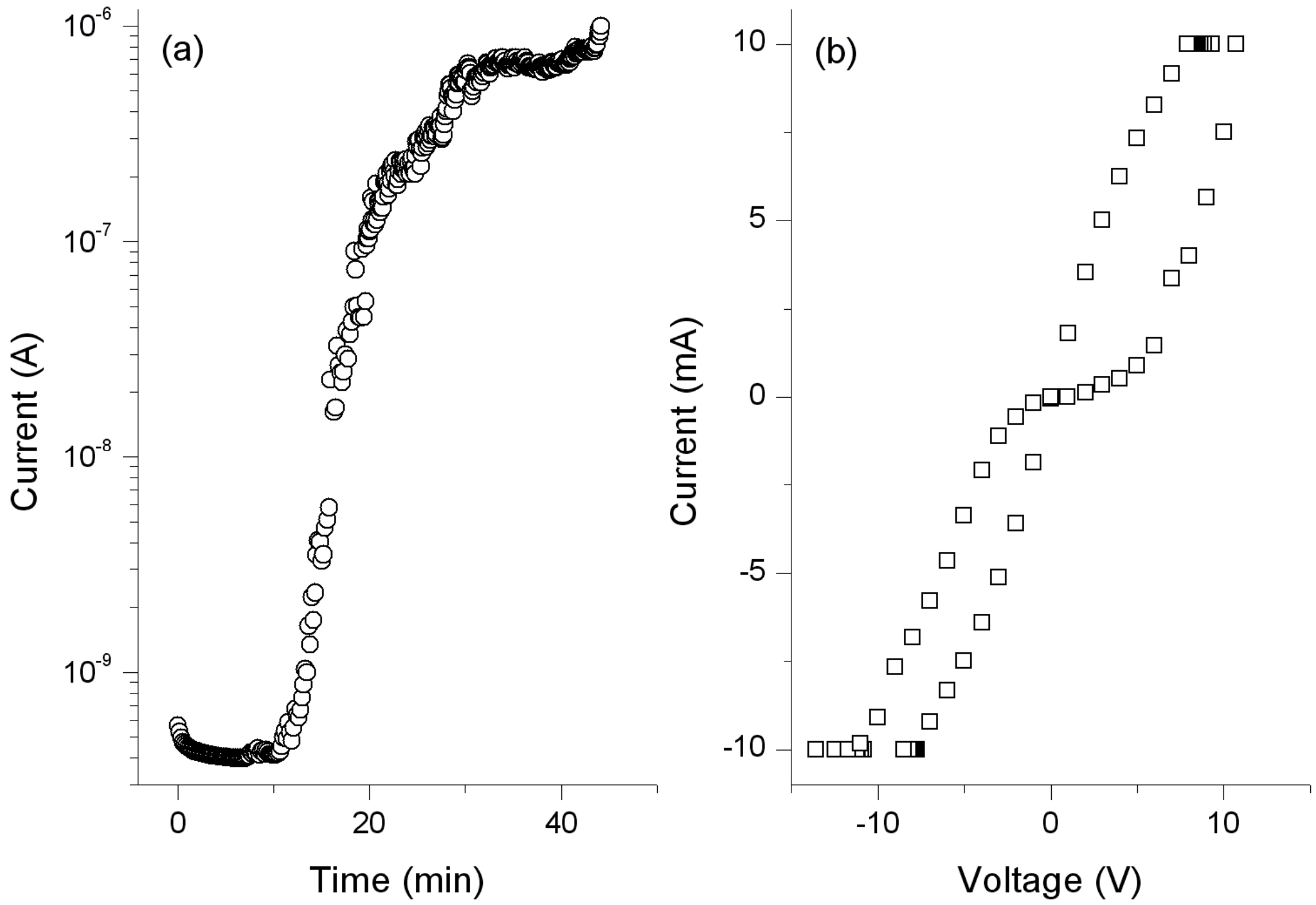

(a)
Current (A)
$10^{-6}$
$10^{-7}$
$10^{-8}$
$10^{-9}$
0
20
40
Time (min)
(b)
Current (mA)
10
5
0
-5
-10
-10
0
10
Voltage (V)

Figure 2

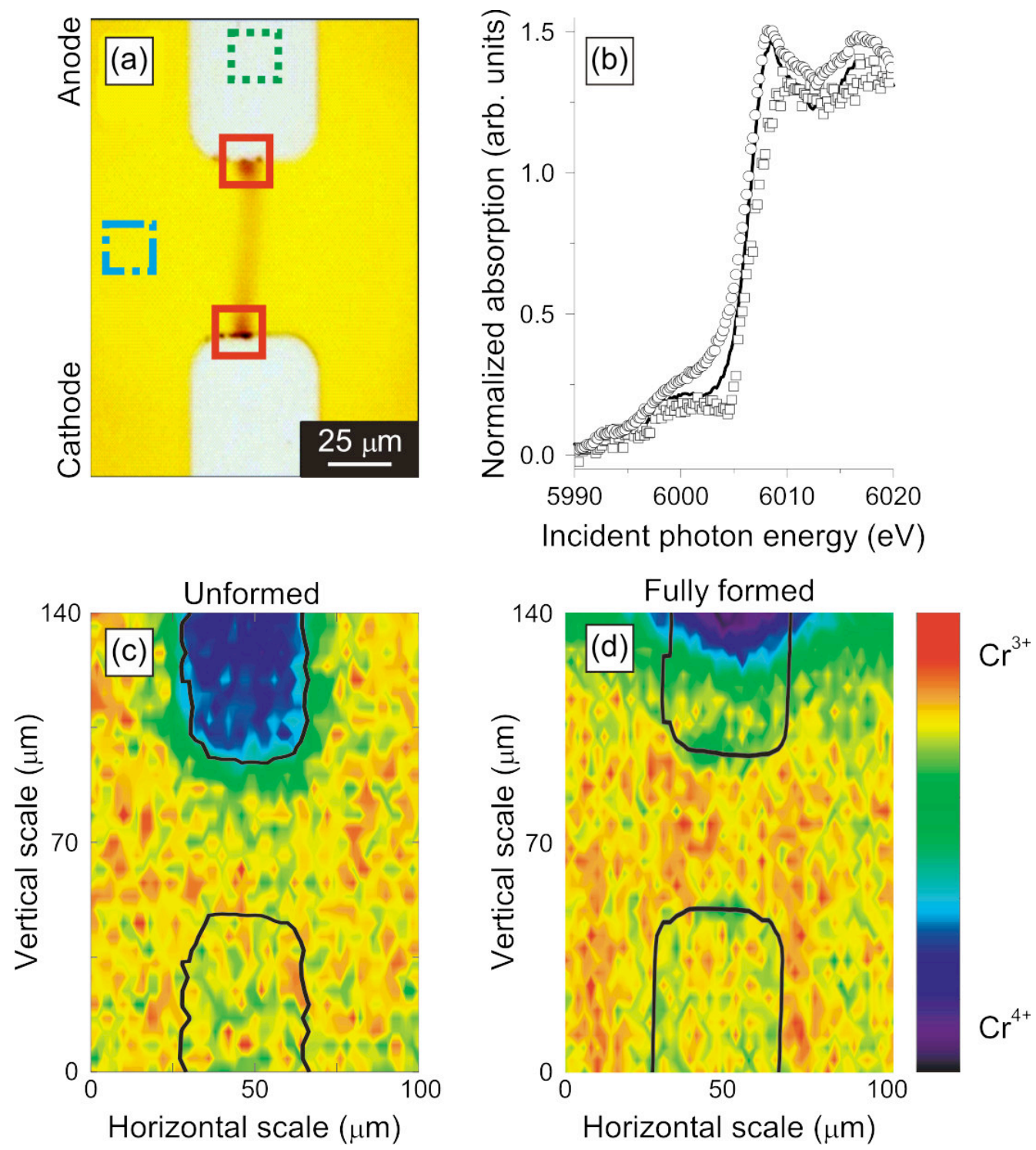

(a)
Anode
Cathode
25 μm
(b)
Normalized absorption (arb. units)
1.5
1.0
0.5
0.0
5990
6000
6010
6020
Incident photon energy (eV)
Unformed
Fully formed
(c)
(d)
140
70
0
Vertical scale (μm)
0
50
100
Horizontal scale (μm)
Cr3+
Cr4+

Figure 3

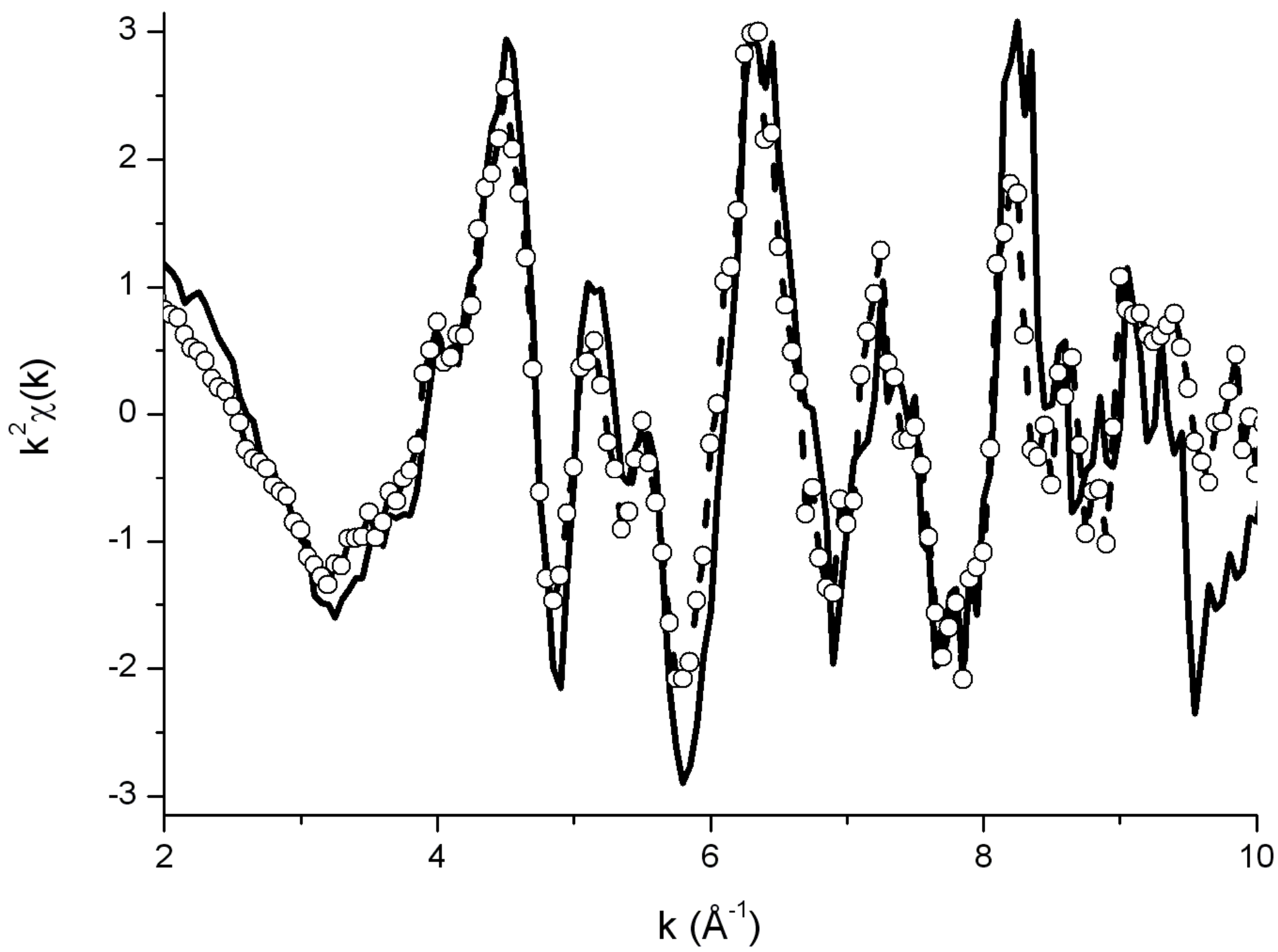

Figure 4

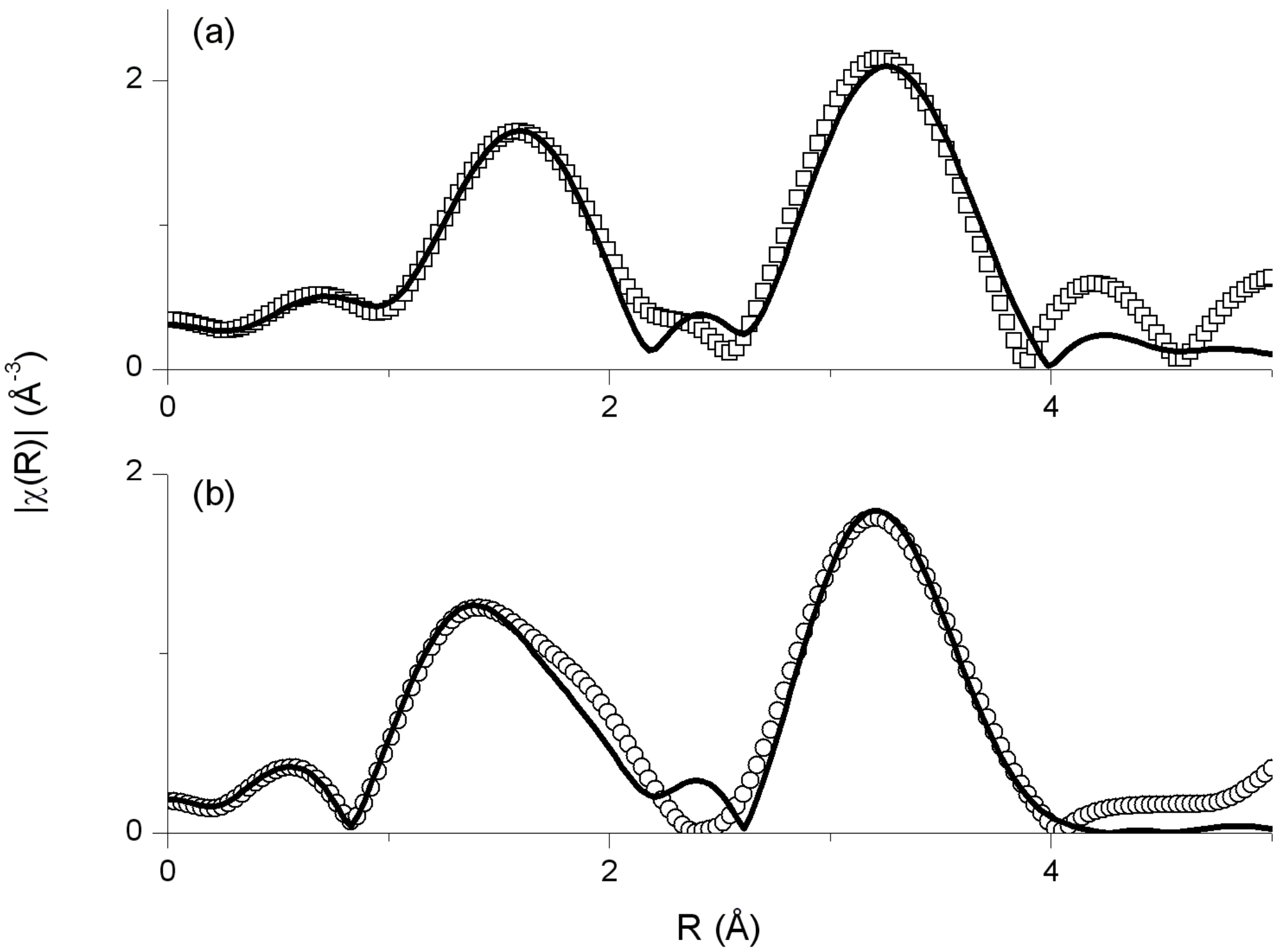